# Remeshing-Free Graph-Based Finite Element Method for Ductile and Brittle Fracture

Avirup Mandal, Parag Chaudhuri and Subhasis Chaudhuri, *Fellow*, IEEE

**Abstract**—Fracture produces new mesh fragments that introduce additional degrees of freedom in the system dynamics. Existing finite element method (FEM) based solutions suffer from an explosion in computational cost as the system matrix size increases. We solve this problem by presenting a graph-based FEM model for fracture simulation that is remeshing-free and easily scales to high-resolution meshes. Our algorithm models fracture on the graph induced in a volumetric mesh with tetrahedral elements. We relabel the edges of the graph using a computed damage variable to initialize and propagate fracture. We prove that non-linear, hyper-elastic strain energy is expressible entirely in terms of the edge lengths of the induced graph. This allows us to reformulate the system dynamics for the relabeled graph without changing the size of system dynamics matrix and thus prevents the computational cost from blowing up. The fractured surface has to be reconstructed explicitly only for visualization purposes. We simulate standard laboratory experiments from structural mechanics and compare the results with corresponding real-world experiments. We fracture objects made of a variety of brittle and ductile materials, and show that our technique offers stability and speed that is unmatched in current literature.

## 1 INTRODUCTION

FRACTURE of objects is an ubiquitous phenomena in the real world. Everyday we come across several instances of fracture, ranging from shattering of brittle glass tumbler to tearing of a soft loaf of bread. In order to recreate such real world phenomena, realistic dynamic fracture simulations are used frequently in computer graphics. Simulating intricate fracture patterns and crack propagation often requires complex mathematical formulation. It has thus been an important research topic in both structural mechanics and computer graphics. Incorporating both brittle and ductile fracture in a single framework needs even more sophisticated analysis.

However, despite intensive research for last two decades, the existing fracture techniques still suffer from scalability issues, poor stability and high computational cost. E.g., as the number of cracks grow in an object, the Finite Element Method (FEM) based approaches proposed in [1] and [2] require remeshing to address the additional degrees of freedom (DOF). This poses challenges like generation of thin (or sliver) elements, instability and heavy computational cost. To avoid these effects authors have explored remeshing-free techniques like eXtended Finite Element Method (XFEM) [3] [4] and the Material Point Method (MPM) [5] [6] [7] [8] [9]. The system matrix of XFEM scales rapidly with the introduction of cracks. Each new crack introduces new DOF, which in turn leads to severe stability issues. On the other hand, the particle based material point methods have shortcomings with respect to imposing boundary conditions, high computational cost and rigidity of fragmented segments. Contrary to the discrete formulations, peridynamics based methods [10] [11] and boundary element methods (BEM) [12] [13] deal with solving the integral equations of continuum mechanics, thus bypassing the difficulty of imposing boundary conditions. But solving these singular integrals is extremely expensive and therefore, is not suitable for low resource applications.

We perform our simulations on volumetric meshes with tetrahedral elements. This underlying mesh, on which all computation is performed, is called the computational mesh. This mesh induces a graph where its vertices become the nodes of the graph and the edges of the mesh become the edges of the graph. The computational mesh is never remeshed. This implies that unlike XFEM, the number of degrees of freedom of the system does not change with the introduction of cracks, i.e., the size of initial system matrix does not change. This allows our method to scale to high resolution models with very little extra computational overhead. The mesh that is rendered for visualization is the same as the computational mesh initially. This mesh however, needs to be split and the fracture surfaces have to be reconstructed for rendering the fracture. This does not affect the computational mesh.

It has been shown in earlier work that the nodal forces of a discretized hyper-elastic FEM system can be completely represented in terms of a function of strain energy density along the edges of the induced graph [14]. We use this fact and follow prior work in structural mechanics [15] to define a purely edge based damage variable, which describes the extent of damage for any edge in the graph. We build over prior work to reformulate the strain energy of damaged elements in 3D volumetric object meshes and then simulate independent movement of the fractured pieces. Further in our model, we can generate and control the diffusion of cracks into the object by imposing a non-local fracture criteria.

- A. Mandal and S. Chaudhuri are with the Department of Electrical Engineering, Indian Institute of Technology Bombay, India.
  Email: avirupmandal@ee.iitb.ac.in, sc@ee.iitb.ac.in.
- P. Chaudhuri is with the Department of Computer Science Engineering, Indian Institute of Technology Bombay, India.
  Email: paragc@sce.iitb.ac.in.

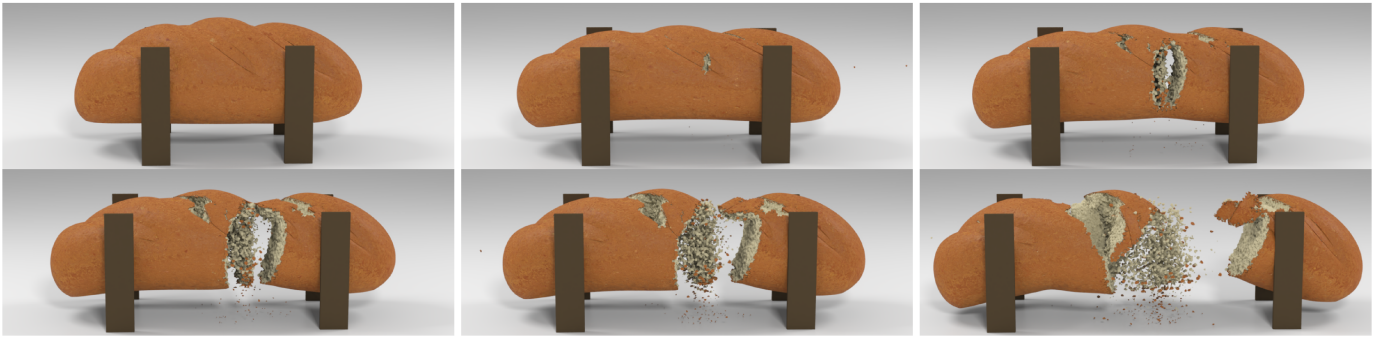

Fig. 1: Our model produces the intricate fracture patterns that result from the tearing of a loaf of bread. We show rendered frames from the simulation (to be seen left to right, top to bottom). The loaf model has around 620k tetrahedra.

We demonstrate the efficacy and robustness of our method by simulating realistic examples of both brittle and ductile fracture. Our method's ability to handle fracture of a variety of materials ranging from stiff glass to squishy jello makes it useful for wide range of applications.

Our contributions are summarized as follows:

- We propose a novel remeshing-free, graph-based, computationally efficient and robust FEM solution for fracture simulation of 3D models.
- We present a generalized version of graph-based FEM algorithm that can simulate fracture of materials with both linear and non-linear strain energy.
- We theoretically prove that hyper-elastic strain energy can be represented in closed form as a function of the length of the edges of any object mesh.
- We extensively validate the correctness of our model against real world structural mechanics experiments.
- We demonstrate via examples that our method serves as a unified FEM framework for simulating both brittle and ductile fracture.

The rest of the paper is organized as follows. We start by presenting a discussion of related works in Section 2. Subsequently, in Section 3 we first describe the theoretical formulation of graph-based FEM, and then detail the technical aspects of fracture via crack initiation and crack propagation. We discuss the implementation details about the visualization of fractured material, collision handling and the complete algorithm for our model in Section 4. Next, we present results for various kinds of materials undergoing fracture, generated using our method in Section 5. In order to validate our method and check its correctness, we compare our results with real world benchmark fracture experiments and existing fracture simulation techniques in literature. Finally, we conclude our paper by discussing the limitations of our work and future work.

## 2 Related Work

We first delve into existing methods for mesh based fracture simulation in computer graphics. Next, we explore meshless particle based fracture simulation methods. We conclude the section by reviewing graph-based finite element analysis presented in material science literature.

### 2.1 Mesh-based Fracture Simulation

The inception of fracture simulation in computer graphics goes back to the seminal work on visco-elastic fracture by Terzopoulos et al. [16]. Early approaches propose to model brittle fracture by removing the springs in a mass-spring system [17] [18] [19] [20] depending on a stress-based yield threshold. But in such a system, where point masses are connected by springs, sudden removal of springs leads to significant visual artifacts. Moreover, visualization of the crack surface often requires the use of a tetrahedral marching algorithm and is therefore computationally expensive [19] [20]. Later, finite element method based solutions for simulating fracture have been more successful and different variations of those approaches are still widely used. The first breakthrough in FEM based fracture came in the paper by O'Brien and Hodgins [2]. In their work, the authors present a nodal stress-based analysis for brittle fracture that was later extended further to ductile fracture [1] [21]. Bao et al. [22] also present a method to simulate both brittle and ductile fracture. These FEM based fracture techniques rely on splitting the elements of a tetrahedral mesh, which satisfy the fracture conditions and then remesh the entire fractured mesh with the crack opening. However, these approaches pose various geometric difficulties like remeshing near crack tips, generation of degenerate elements, forming ill-conditioned basis matrices. Subsequent FEM based approaches employ different techniques to alleviate these problems, as such, dynamic local mesh refinement to repair degenerate tetrahedra [23], remeshing depending on gradient descent flow for finer fracture resolution [24], adaptive subdivision schemes for tetrahedral [25] and triangular meshes [26] where remeshing is done around the high stress areas to improve fracture resolution.

FEM based solutions that work with a virtual node algorithm (VNA) [27] duplicate elements and add extra degrees of freedom to facilitate partial or full crack openings, instead of splitting the mesh elements. Later, VNA based methods are improved to incorporate cutting at lower than mesh resolution [28] and to robustly handle intersections passing through node, edge or face through an adaptive element duplication approach [29]. Recently, Koschier et al. [3] presented a resmeshing-free extended finite element method (XFEM) based algorithm to simulate the dynamic pre-defined cuts in a 3D mesh. Like VNA, XFEM too adds extra degrees of freedom using enrichment functions but the

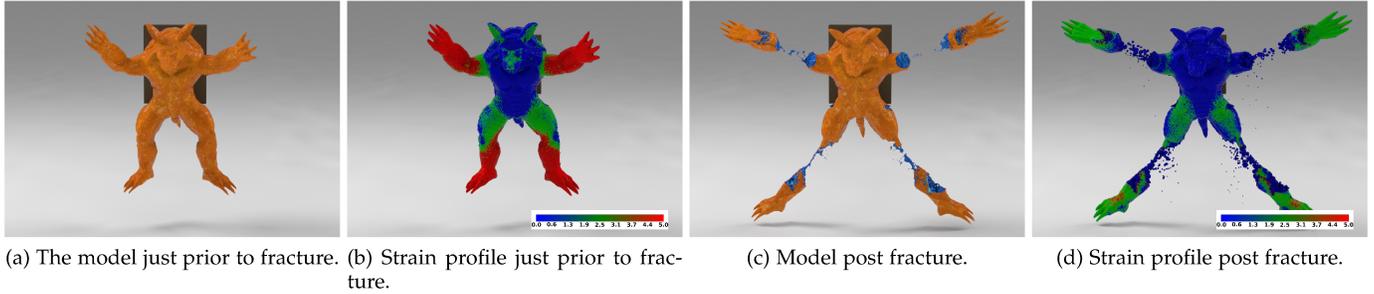

(a) The model just prior to fracture. (b) Strain profile just prior to fracture. (c) Model post fracture. (d) Strain profile post fracture.

Fig. 2: Here we show frames from a simulation where we rip off the limbs of a jello armadillo. In the strain profiles, red to blue colour gradient denotes highest to lowest strain values. The model fractures where the strain is highest prior to fracture. Post fracture, the strain drops.

advantage of XFEM over VNA is while VNA duplicates the mesh element leading to an erratic conservation of mass, XFEM splits the initial mass of an element into fragments. Chitalu et al. [4] expanded the pre-defined cuts to crack generation and propagation, and also incorporated crack-tip enrichment for simulating partially fractured tetrahedra.

Contrary to processing fracture in volume elements like FEM, boundary element methods (BEM) simulate cracks on a surface mesh. Even though BEM found its way into graphics by James et al. [30] for simulating elastostatic deformable objects, it has been used recently for fracture simulation. Hahn et al. [12] and Zhu et al. [31] used BEM with stress intensity factors (SIFs) along crack fronts to successfully simulate brittle fracture. Later Hahn et al. [13] expanded their work to develop an algorithm for fast approximation of BEM brittle fracture, alleviating the high computational cost for solving singular integrals. Currently, to the best of our knowledge, there exists no work on BEM ductile fracture in graphics.

Peridynamics methods solve the integral equations of continuum mechanics. Unlike discrete differential-based methods, e.g. FEM or MPM, which need to impose necessary boundary conditions for simulating fracture, it is trivial to model discontinuities using integral-based peridynamics. In initial work using these ideas, authors [11] approximate the behaviour of peridynamics using mass-spring systems to simulate brittle fracture. Later work [10] presents original peridynamics based fracture algorithms to simulate brittle as well as ductile fracture.

### 2.2 Meshless Fracture Simulation

Among the meshless methods, material point methods have gained considerable recognition in recent years. First major breakthrough of MPM in computer graphics came through the seminal work in snow simulation by Stomakhin et al. [32]. Later MPM has found its application in simulating a wide range of materials as such chocolate, lava, hybrid fluids and rubber to name a few. Interested readers can look at [33] for a detailed overview of different MPM methods. Most recently, MPM has shown promising results in simulating fracture due to its ability to handle extreme topological change. Hu et al. [7] propose the Compatible Particle-in-Cell (CPIC) algorithm, which allows the simulation of sharp discontinuities inside a material and thus simulates dynamic material cutting. Later Wolper et al. [8], introduce

Continuum Damage Mechanics (CDM) with a variational energy-based formulation for crack evolution. This achieves a realistic dynamic fracture simulation for isotropic materials. Further study [9] devoted to formulating non-local CDM to simulate anisotropic material fracture.

MPM methods reproduce realistic fracture phenomena but they are not well suited to rendering sharp boundaries. They not only have difficulty in enforcing essential boundary conditions, they also suffer from disappearance of intricate crack patters due to excessive smoothing, high computational cost and rigidity of fragmented elements. Further, MPM methods struggle to simulate rigid object fracture.

### 2.3 Graph-based Finite Element Method

While simulating fractures in the material domain, classical FEM introduces degenerate elements due to remeshing and XFEM has severe limitations to incorporating a large number of cracks due to rapid scaling up of the system matrix. Moreover, both of them suffer from stability issues and high computational cost. While BEM and peridynamics overcome these problems by solving direct integrals, they demand prohibitively high numerical computation. Again, use of BEM is limited to materials with large volumes. Meshless methods have their own limitations as described above.

Work in finite element analysis literature by Reddy et al. [14] proposes a solution based on classical FEM to show that for any hyper-elasticity problem, the magnitude of the nodal forces can be written completely in terms of strain energy density along the edges composing the elements, while force directions are along the unit vectors corresponding to the edges. Using this idea in subsequent work, authors [15] introduce Graph-based Finite Element Analysis (Gra-FEA) where a damage variable corresponding to the edges of composing elements is used. It is based on the strain threshold and thus can simulate the fracture of an object by weakening the material. The advantage of using Gra-FEA is that it requires no remeshing, while adding little computational overhead on FEM and it still retains all the advantages of FEM. The dependency of Gra-FEA on the underlying mesh structure is thoroughly studied in another work [34]. Our work introduces Gra-FEA to the visual simulation of fracture for computer graphics. Existing literature referred above on Gra-FEA only focuses on 2D materials consisting of triangular elements with linear strain energy. We extend the idea to 3D domain and our fracture algorithm can handle linear as well as non-linear strain energy. The original Gra-FEA formulation is restricted to simulating fracture for brittle materials. Our algorithm is capable of simulating fracture for both brittle and ductile materials. We believe, to the best of our knowledge, that this is the first work in computer graphics to simulate fracture of 3D brittle, as well as ductile objects, with linear or non-linear strain energy using the graph-based FEM.

## 3 Governing Methods

We briefly derive the formulation of graph-based FEM from the theory of classical FEM. Then we delve into the details of fracture generation using graph-based FEM. In our work, we always use tetrahedral finite elements with a linear basis for domain discretization.

### 3.1 Classical FEM

Consider a three-dimensional domain $\Omega \in \mathbb{R}^3$ that is discretized into a mesh of $n_{tet}$ tetrahedra. Moreover, let $\mathfrak{I}$ be the set of nodes $n_v$ shared by the tetrahedral mesh elements $\Delta_e$. As shown in Figure 3, a displacement function, $\mathbf{u}: \Omega \times [0, \infty) \longrightarrow \mathbb{R}^3$, can be defined as a mapping from a material point $\xi \in \Omega$ to its deformed location $\mathbf{x} \in \Omega_t$ in the world space. $\Omega_t$ represents the world space at time $t$. At a certain timestep $t \in [0, \infty)$, the displacement function can be represented as

$$\mathbf{u}(\xi, t) = \sum_{i \in \mathfrak{I}} \mathbf{N}_i(\xi) \mathbf{u}_i(t), \quad (1)$$

where $\mathbf{N}_i(\xi)$ is the shape function and $\mathbf{u}_i$ is the displacement vector at the node $i$. For the sake of brevity of notation, we drop the shape ($\xi$) and time ($t$) parameters for the shape and displacement functions in further discussion.

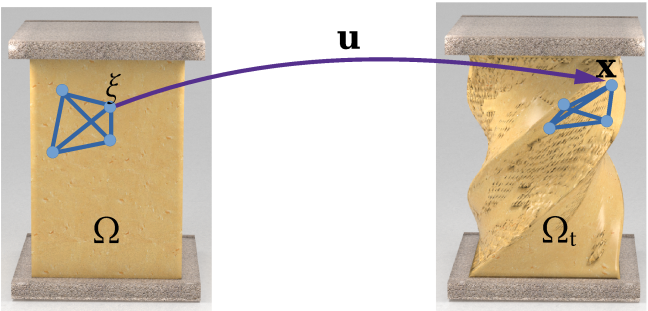

Fig. 3: Deformable object: A displacement function maps a point on the object in material coordinates on the left and to world coordinates on the right.

The core idea of deformable object simulation is that when an object mesh is deformed from its rest position, a hyper-elastic strain energy $\Psi$, develops inside it. The strain energy produces internal elastic force $\mathbf{f}^{int}$, which in turn tries to restore the rest shape of the mesh. Thus, the system dynamics of deformed object can be written in Lagrange's form as

$$\mathbf{M}\ddot{\bar{\mathbf{u}}} + \mathbf{B}\dot{\bar{\mathbf{u}}} + \mathbf{f}^{int} = \mathbf{f}^{ext} \quad (2)$$

$$\bar{\mathbf{u}} = \left(\mathbf{u}_1^T \ldots \mathbf{u}_{n_v}^T\right)^T \quad (3)$$

Here $\mathbf{M}$ and $\mathbf{B}$ are the mass and damping matrices of full system. $\mathbf{f}^{ext}$ and $\mathbf{f}^{int}$ represent the external force vector and internal force vector respectively. Now Equation 2 can further be simplified as the sum of parameters from each individual tetrahedral element.

$$\sum_{e=1}^{n_{tet}} \mathbf{m}_e \ddot{\bar{\mathbf{u}}} + \sum_{e=1}^{n_{tet}} \mathbf{b}_e \dot{\bar{\mathbf{u}}} + \sum_{e=1}^{n_{tet}} \mathbf{f}_e^{int} = \sum_{e=1}^{n_{tet}} \mathbf{f}_e^{ext} \quad (4)$$

$$\mathbf{m}_e = \int_{\Delta_e} \rho_0 \mathbf{N}_e^T \mathbf{N}_e d\xi \quad (5)$$

$$\mathbf{f}_e^{ext} = \int_{\Delta_e} \mathbf{N}_e^T \mathbf{t}_e d\xi \quad (6)$$

$$\mathbf{f}_e^{int} = \int_{\Delta_e} \mathbf{s}_e(\bar{\mathbf{u}}, \xi) d\xi \quad (7)$$

$$\mathbf{N}_e = [\mathbf{N}_0 \mathbf{I}_3 \ \mathbf{N}_1 \mathbf{I}_3 \ \mathbf{N}_2 \mathbf{I}_3 \ \mathbf{N}_3 \mathbf{I}_3]. \quad (8)$$

While $\mathbf{m}_e$ is the mass matrix of a tetrahedral element, $\mathbf{N}_e$, $\mathbf{t}_e$, $\mathbf{f}_e^{ext}$ and $\mathbf{f}_e^{int}$ represent element shape function vector, reference body force, external element force vector and internal element force vector respectively. $\mathbf{I}_3$ refers to $3 \times 3$ identity matrix.

Specific elastic element force, $\mathbf{s}_e$, is defined as

$$\mathbf{s}_e = \frac{\partial \mathbf{\Psi}^e}{\partial \mathbf{u}} \quad (9)$$

where $\mathbf{\Psi}^e$ is elemental hyper-elastic strain energy.

$\mathbf{b}_e = \alpha \mathbf{m}_e + \beta \mathbf{k}_e + \tilde{\mathbf{b}}_e$ is the per element control damping matrix. Scalar quantities $\alpha$ and $\beta$ control the level of 'mass' and 'stiffness' damping respectively while $\tilde{\mathbf{b}}_e$ is a user defined additional damping matrix.

The gradient of $\mathbf{f}_e^{int}$ with respect to $\mathbf{u}$ is called tangent stiffness matrix $\mathbf{k}_e$.

$$\mathbf{k}_e = \int_{\Delta_e} \frac{\partial \mathbf{f}_e^{int}}{\partial \mathbf{u}} d\xi \quad (10)$$

The stiffness matrix of full system, $\mathbf{K}$, can be represented as

$$\mathbf{K} = \sum_{e=1}^{n_{tet}} \mathbf{k}_e \quad (11)$$

Interested readers can look into the review paper by Sifakis and Barbič [35] for further details. In this work, we developed our fracture simulation model based on works by Smith et al. [36], Sin et al. [37] and Bargteil et al. [38]. Table 1 summarizes some of the quantities that we use frequently in FEM simulations.

### 3.2 Graph-based FEM

Let the hyper-elastic strain energy of a tetrahedral element, $\Delta_e$, be $\mathbf{\Psi}^e$ and $\mathbf{f}_{e_i}^{int}$ be the internal elastic force acting on the $i^{th}$ node of the element. Now it can be proved [14] that nodal internal elastic forces of $\Delta_e$ can always be decomposed along the vectors connecting the nodes of $\Delta_e$. The statement can be mathematically formulated as

$$\mathbf{f}_{e_i}^{int} = -V_e \frac{\partial \mathbf{\Psi}^e}{\partial \mathbf{r}_i^e} = 2V_e \sum_{\substack{j=1 \\ j \neq i}}^{n_e} \frac{\partial \mathbf{\Psi}^e}{\partial \left(d_{ij}^e\right)^2} d_{ij}^e \hat{\mathbf{d}}_{ij}^e \quad (12)$$

| Symbol | Definition |
| --- | --- |
| $\mathbf{F} = \mathbf{I} + \nabla_\xi \mathbf{u}$ | Deformation gradient |
| $J = \det(\mathbf{F})$ | Relative volume change |
| $\mathbf{C} = \mathbf{F}^T \mathbf{F}$ | Right Cauchy-Green deformation tensor |
| $I_C = \text{trace}(\mathbf{C})$ | First Right Cauchy-Green invariant |
| $II_C = \mathbf{C} : \mathbf{C}$ | Second Right Cauchy-Green invariant |
| $III_C = \det(\mathbf{C})$ | Third Right Cauchy-Green invariant |
| $\mathbf{E} = \frac{1}{2}\left(\mathbf{F}^T\mathbf{F} - 1\right)$ | Lagrangian finite strain tensor |
| $\mathbf{P}(\mathbf{F}) = \frac{\partial \mathbf{\Psi}}{\partial \mathbf{F}}$ | First Piola-Kirchhoff stress tensor |

TABLE 1: Quantities frequently used in FEM

where $d_{ij}^e = ||\mathbf{r}_i^e - \mathbf{r}_j^e||$ is the distance between the $i^{th}$ and $j^{th}$ nodes of $\Delta_e$ and $\hat{\mathbf{d}}_{ij}^e$ is unit vector along $d_{ij}^e$. Moreover, $V_e$ and $n_e$ represent volume and the number of nodes of $\Delta_e$ respectively. It can be seen from Equation 12 that the magnitude of the nodal internal elastic forces depend only on strain energy density of the element while the force directions are along its edges.

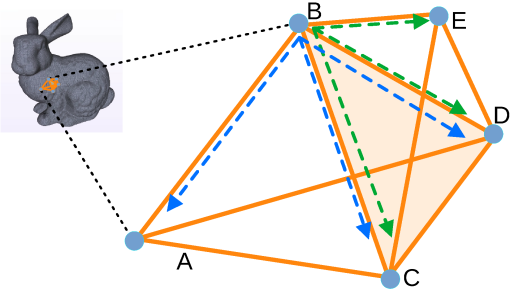

Fig. 4: Nodal force distribution along the edges for graph-based FEM.

Figure 4 shows two tetrahedral elements, $\Delta_e ABDC$ and $\Delta_e BEDC$, which share a common face. These can be thought of as being a part of a larger object mesh. We assume that the overall mesh is deformed and each node in the mesh develops internal elastic force. Now according to Equation 12, internal elastic force on the node $B$ for $\Delta_e ABDC$ can be written along the edges of $\Delta_e ABDC$ as $\mathbf{f}_{\Delta_e ABDC}^B = \mathbf{f}_{\Delta_e ABDC}^{BC} + \mathbf{f}_{\Delta_e ABDC}^{BA} + \mathbf{f}_{\Delta_e ABDC}^{BD}$ (shown as dotted blue lines). Similarly, internal elastic force on the node $B$ for $\Delta_e BEDC$ can be written along the edges of $\Delta_e BEDC$ as $\mathbf{f}_{\Delta_e BEDC}^B = \mathbf{f}_{\Delta_e BEDC}^{BC} + \mathbf{f}_{\Delta_e BEDC}^{BD} + \mathbf{f}_{\Delta_e BEDC}^{BE}$ (shown as dotted green lines). The distribution of nodal internal elastic forces for the entire mesh can be computed similarly.

### 3.3 Fracture with Graph-based FEM

In this section, we present our model for edge based fracture derived from Graph-based FEM. We take advantage of the fact that stress of an element is always a function of hyper-elastic strain energy density. We can find normal stress in the direction of the edges of a tetrahedral element by using a suitable transformation. Next, depending on the state of edge i.e., broken or unbroken, we manipulate the normal stress along the edges to degrade the strain energy density for a damaged element. Finally, using this updated strain energy density, we recalculate the elastic forces and tangent stiffness matrix for fracture simulation.

Using chain rule of differentiation and deformation gradient from Table 1, we can rewrite the Equation 12 as

$$\mathbf{f}_{e_i}^{int} = 2V_e \sum_{\substack{j=1 \\ j \neq i}}^{n_e} \sum_{k=1}^{3} \sum_{l=1}^{3} \left[ \frac{\partial \mathbf{\Psi}^e}{\partial [\mathbf{F}]_{kl}} \frac{\partial [\mathbf{F}]_{kl}}{\partial \left(d_{ij}^e\right)^2} \right] d_{ij}^e \hat{\mathbf{d}}_{ij}^e \quad (13)$$

where $\frac{\partial \mathbf{\Psi}^e}{\partial [\mathbf{F}]_{kl}}$ are components of first Piola-Kirchhoff stress tensor of $\Delta_e$ (see Table 1). Thus, the edge based elastic forces can be completely represented as a function of stress in $\Delta_e$. This stress in turn depends on the derivative of element strain energy, $\mathbf{\Psi}^e$ w.r.t. the deformation gradient $\mathbf{F}$. Therefore, by manipulating the stress components properly we can change the internal elastic forces and in turn simulate fracture of an element.

Let the rectangular Cartesian components of the Piola-Kirchhoff stress tensor, $\boldsymbol{\sigma}_c^e$, be denoted as

$$\boldsymbol{\sigma}_c^e = \begin{bmatrix} \sigma_{xx}^e & \sigma_{yy}^e & \sigma_{zz}^e & \sigma_{xy}^e & \sigma_{xz}^e & \sigma_{yz}^e \end{bmatrix} \quad (14)$$

The set of normal stress along the edges is represented as

$$\boldsymbol{\sigma}_f^e = \begin{bmatrix} \sigma_{12}^e & \sigma_{13}^e & \sigma_{14}^e & \sigma_{23}^e & \sigma_{24}^e & \sigma_{34}^e \end{bmatrix} \quad (15)$$

where $\sigma_{mn}^e$ represents the normal stress along the edge formed by nodes $m$ and $n$ in Equation 15. Then the transformation of Cartesian stress to normal stress in the direction of an edge can be formulated as below [14] [15]

$$\sigma_{mn}^e = \sigma_{xx}^e \cos^2 \phi_x + \sigma_{yy}^e \cos^2 \phi_y + \sigma_{zz}^e \cos^2 \phi_z \\ + \sigma_{xy}^e \cos \phi_x \cos \phi_y + \sigma_{xz}^e \cos \phi_x \cos \phi_z + \sigma_{yz}^e \cos \phi_y \cos \phi_z \quad (16)$$

Assuming that $\hat{\mathbf{d}}_{mn}^e, \hat{\mathbf{x}}, \hat{\mathbf{y}}$ and $\hat{\mathbf{z}}$ denote unit vectors along the edge formed by nodes $m$ and $n$ of the tetrahedral element $\Delta_e$, $x$-axis, $y$-axis and $z$-axis respectively, the following relations hold true

$$\cos \phi_x = \hat{\mathbf{d}}_{mn}^e \cdot \hat{\mathbf{x}}, \quad \cos \phi_y = \hat{\mathbf{d}}_{mn}^e \cdot \hat{\mathbf{y}}, \quad \cos \phi_z = \hat{\mathbf{d}}_{mn}^e \cdot \hat{\mathbf{z}}, \quad (17)$$

Using Equation 16, we can write the final relation between two sets of strains as:

$$\begin{bmatrix} \sigma_{12}^e & \sigma_{13}^e & \sigma_{14}^e & \sigma_{23}^e & \sigma_{24}^e & \sigma_{34}^e \end{bmatrix}^T = \mathbf{T} \begin{bmatrix} \sigma_{xx}^e \\ \sigma_{yy}^e \\ \sigma_{zz}^e \\ \sigma_{xy}^e \\ \sigma_{xz}^e \\ \sigma_{yz}^e \end{bmatrix} \quad (18)$$

where, for convenience of notation, if we denote $\cos \phi$ as $c\phi$, then

$$\mathbf{T} = \begin{bmatrix} c^2\phi_x^{12} & c^2\phi_y^{12} & c^2\phi_z^{12} & c\phi_x^{12}c\phi_y^{12} & c\phi_x^{12}c\phi_z^{12} & c\phi_y^{12}c\phi_z^{12} \\ c^2\phi_x^{13} & c^2\phi_y^{13} & c^2\phi_z^{13} & c\phi_x^{13}c\phi_y^{13} & c\phi_x^{13}c\phi_z^{13} & c\phi_y^{13}c\phi_z^{13} \\ c^2\phi_x^{14} & c^2\phi_y^{14} & c^2\phi_z^{14} & c\phi_x^{14}c\phi_y^{14} & c\phi_x^{14}c\phi_z^{14} & c\phi_y^{14}c\phi_z^{14} \\ c^2\phi_x^{23} & c^2\phi_y^{23} & c^2\phi_z^{23} & c\phi_x^{23}c\phi_y^{23} & c\phi_x^{23}c\phi_z^{23} & c\phi_y^{23}c\phi_z^{23} \\ c^2\phi_x^{24} & c^2\phi_y^{24} & c^2\phi_z^{24} & c\phi_x^{24}c\phi_y^{24} & c\phi_x^{24}c\phi_z^{24} & c\phi_y^{24}c\phi_z^{24} \\ c^2\phi_x^{34} & c^2\phi_y^{34} & c^2\phi_z^{34} & c\phi_x^{34}c\phi_y^{34} & c\phi_x^{34}c\phi_z^{34} & c\phi_y^{34}c\phi_z^{34} \end{bmatrix} \quad (19)$$

The criteria for fracture in graph-based FEM is that if the weighted average of normal stress along the direction of any edge exceeds the critical stress threshold $\sigma_{thres}^e$, a crack is to be formed on that edge. Using Equations 14 and 18 the fracture criteria can, therefore, be defined as

$$\sigma_{mn}^{e^*} = \left[ \mathbf{T}\left( \sum_{||\mathbf{r}-\mathbf{r}_0|| \leq R_d} \omega(\mathbf{r}-\mathbf{r}_0)\boldsymbol{\sigma}_c^e(\mathbf{r}) \right)\right]_{mn} \geq \sigma_{thres}^e \quad (20)$$

where $\sigma_{mn}^{e^*}$ is a component of weighted average of normal stress along the edge formed by nodes $m$ and $n$, $\omega$ is a weight kernel and $R_d$ is support of the kernel. While vector $\mathbf{r}$ refers to position of centroid of any tetrahedron in the mesh, $\mathbf{r}_0$ denotes the position of centroid of the reference tetrahedron. As the support of kernel, $R_d$, increases, the cracks become more diffused i.e., they spread more inside the object [34]. For $R_d = 0$, the fracture criteria depends only on the reference tetrahedron, resembling a local fracture. Now for each edge of the mesh, we maintain a binary damage variable, $\zeta \in \{0, 1\}$. A value of $\zeta = 0$ corresponds to unbroken edge, while $\zeta = 1$ implies crack is formed on the edge.

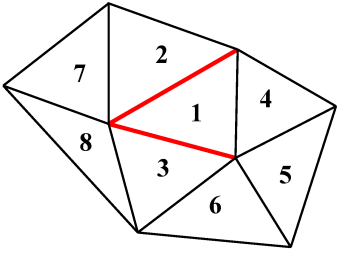

Fig. 5: Fractured edges shared multiple elements.

$$\zeta = \begin{cases} 0 & \text{if } \sigma_{mn}^{e^*} < \sigma_{thres} \\ 1 & \text{if } \sigma_{mn}^{e^*} \geq \sigma_{thres} \end{cases} \quad (21)$$

As evident from Equation 20, the number of broken edges in a tetrahedral element may vary from zero to six. Once an edge is broken, it will not get repaired in subsequent iterations and all the tetrahedra that share the broken edge will be considered damaged. As an example, a shared edge fracture is depicted in Figure 5, using a simplified, two-dimensional example. Two edges in the mesh, colored red, are broken. These broken edges are shared by triangles 1 (two edges), 2 (one edge) and 3 (one edge). In the subsequent iterations, all three triangular elements will be considered damaged with varying degrees of damage. Whenever an edge, $e_{br}$, gets broken, the corresponding normal stress of that edge becomes zero for all elements sharing that edge. Using Equations 18 and 21, we can write

$$\begin{bmatrix} \sigma_{12}^{e'} \\ \sigma_{13}^{e'} \\ \sigma_{14}^{e'} \\ \sigma_{23}^{e'} \\ \sigma_{24}^{e'} \\ \sigma_{34}^{e'} \end{bmatrix} = \begin{bmatrix} \zeta_{12} & 0 & 0 & 0 & 0 & 0 \\ 0 & \zeta_{13} & 0 & 0 & 0 & 0 \\ 0 & 0 & \zeta_{14} & 0 & 0 & 0 \\ 0 & 0 & 0 & \zeta_{23} & 0 & 0 \\ 0 & 0 & 0 & 0 & \zeta_{24} & 0 \\ 0 & 0 & 0 & 0 & 0 & \zeta_{34} \end{bmatrix} \begin{bmatrix} \sigma_{12}^e \\ \sigma_{13}^e \\ \sigma_{14}^e \\ \sigma_{23}^e \\ \sigma_{24}^e \\ \sigma_{34}^e \end{bmatrix} \forall e_{br} \in \Delta_e \quad (22)$$

where $\zeta_{mn}$ and $\sigma_{mn}^{e'}$ are damage variable and stress after fracture along the edge formed by nodes $m$ and $n$. In matrix form Equation 22 is written as

$$\boldsymbol{\sigma}_f^{e'} = \zeta \boldsymbol{\sigma}_f^e \quad (23)$$

### 3.4 Hyper-elastic Strain Energy for Fracture

We update system equation of fractured mesh using a reformulation of elastic energy. Existing work on Gra-FEA [15] defines the updated elastic energy as a function of edge based stress $\boldsymbol{\sigma}_f^{e'}$ such as

$$\boldsymbol{\Psi}_{new}^e = \frac{1}{2}\boldsymbol{\sigma}_f^{e'} \cdot \boldsymbol{\epsilon}_f^{e'} \quad (24)$$

where using transformation matrix from Equation 19, we can write $\boldsymbol{\epsilon}_f^{e'} = \mathbf{T}\mathbf{E}^{-1}\mathbf{T}^{-1}\boldsymbol{\sigma}_f^{e'}$ and $\mathbf{E}$ is Young modulus matrix [21]. However, this kind of update is applicable only for linear elasticity. For non-linear elasticity the elastic energy can not be directly written with $\boldsymbol{\sigma}_f^{e'}$ and $\boldsymbol{\epsilon}_f^{e'}$. Thus this kind of reformulation of elastic energy is not possible for non-linear models.

Our reformulation of elastic energy works for non-linear elasticity. It also makes graph-based amenable for visual simulation and generalizes it to 3D models.

Using Equation 22 the new elastic energy for fractured elements is defined as

$$\begin{aligned}\boldsymbol{\Psi}_{new}^e &= \left( \frac{|\sigma_{12}^{e'}| + |\sigma_{13}^{e'}| + |\sigma_{14}^{e'}| + |\sigma_{23}^{e'}| + |\sigma_{24}^{e'}| + |\sigma_{34}^{e'}|}{|\sigma_{12}^e| + |\sigma_{13}^e| + |\sigma_{14}^e| + |\sigma_{23}^e| + |\sigma_{24}^e| + |\sigma_{34}^e|} \right) \boldsymbol{\Psi}^e \\ &= \chi^e \boldsymbol{\Psi}^e \end{aligned} \quad (25)$$

where $\chi^e$ signifies the fractured elastic strain energy as fraction of original one depending on the number of damaged edges. It is important to note here that when an element fractures into two or more disjoint fragments, we assign zero value to the fractured elastic strain energy i.e. $\boldsymbol{\Psi}_{new}^e = 0$, in Equation 25.

However, for this to work in Equation 12 to hold, we need to show that hyper-elastic strain energy $\boldsymbol{\Psi}^e$ can be completely represented using just $d_{ij}^e$. We prove this in the next section.

### 3.5 Proof for Edge Length Dependence of Strain Energy

Along with the Right Cauchy-Green deformation tensor $\mathbf{C}$ and the three invariants $I_C, II_C$ and $III_C$, as defined in Table 1, let us introduce two more well-known anisotropic invariants [39]

$$\begin{aligned} IV_C &= \mathbf{a}^T\mathbf{C}\mathbf{b} \\ V_C &= \mathbf{a}^T\mathbf{C}^T\mathbf{C}\mathbf{b} \end{aligned} \quad (26)$$

where $\mathbf{a}, \mathbf{b}$ are constant anisotropic fiber directions and they may or may not be equal to each other. Hyper-elastic energy can generally be represented using a subset of these five invariants as $\boldsymbol{\Psi}^e = f(I_C, II_C, III_C, IV_C, V_C)$. Therefore, in order to show that the hyper-elastic strain energy can be expressed as a function of the length of edges of a mesh, it is sufficient to show that every element of the set of invariants can be expressed in closed form using only the length of the edges of the mesh in a graph-based FEM setting.

**Theorem 3.1.** *Every element of the set of invariants* $\mathbf{I}_V = \{I_C, II_C, III_C, IV_C, V_C\}$ *can be expressed in closed form using only the length of the edges of a mesh used in FEM.*

*Proof.* Let us assume an element of a mesh used in FEM, $\boldsymbol{\Xi}_e$, in $k$-dimensional space, which consists of $Z_l$ edges. Let $\lambda_{mn}$

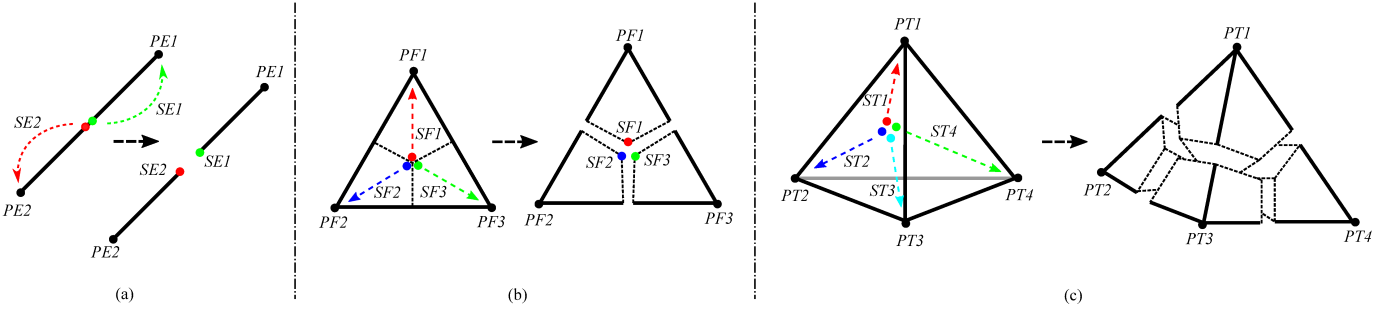

Fig. 6: Splitting of edge (left), face (middle) and tetrahedron (right) for visualization of fracture.

be the stretch ratio of an edge formed by nodes $m$ and $n$ of the same element,

$$\lambda_{mn}^2 = \left(\frac{d_{mn}^e}{D_{mn}^e}\right)^2 \tag{27}$$

where $d_{mn}^e$ and $D_{mn}^e$ denote current and initial length of the edge respectively.

Now, following similar arguments presented in Equation 16 and using the Right Cauchy-Green deformation tensor $\mathbf{C}$, we can also write,

$$\lambda_{mn}^2 = \mathbf{C} \bullet (\mathbf{q}_{mn}^e \otimes \mathbf{q}_{mn}^e) = \mathbf{C} \bullet \mathbf{Q}_{mn}^e \tag{28}$$

where $\otimes$ denotes tensor product and $\mathbf{Q}_{mn}^e \in \mathbb{R}^{k \times k}$. Similar to Equation 17, $\mathbf{q}_{mn}^e$ can be written in a $k$-dimensional space as

$$\mathbf{q}_{mn}^e = \begin{bmatrix} \hat{\mathbf{d}}_{mn}^e \cdot \hat{\mathbf{i}}_1 & \hat{\mathbf{d}}_{mn}^e \cdot \hat{\mathbf{i}}_2 & \dots & \hat{\mathbf{d}}_{mn}^e \cdot \hat{\mathbf{i}}_k \end{bmatrix}^T$$

where $\{\hat{\mathbf{i}}_1, \hat{\mathbf{i}}_2 \dots \hat{\mathbf{i}}_k\}$ denotes a set of orthogonal basis vectors of the $k$-dimensional space.

Extending this formulation to all $Z_l$ edges, we get

$$\mathbf{\Lambda} = \mathbf{C} \bullet \mathbf{Q} \tag{29}$$

where $\mathbf{\Lambda} \in \mathbb{R}^{Z_l}$ and $\mathbf{Q} \in \mathbb{R}^{k \times k \times Z_l}$, whose entries are equal to $\lambda_{mn}^2$ and $\mathbf{Q}_{mn}^e$ respectively.

In order to express $\mathbf{C}$ in terms of $\lambda_{mn}^e$ we need to invert $\mathbf{Q}$, which is not possible. However, we can determine $\mathbf{C}$ as the solution to the following optimization problem

$$\mathcal{L} = \underset{\hat{\mathbf{C}}}{\operatorname{argmin}} \ ||\hat{\mathbf{C}} \bullet \mathbf{Q} - \mathbf{\Lambda}||^2 \tag{30}$$

The solution to this optimization problem allows us to express $\mathbf{C}$ in terms of $\mathbf{\Lambda}$ and $\mathbf{Q}$.

The solution to the problem is given in closed form by

$$\hat{\mathbf{C}} = \mathbf{\Lambda} \bullet [\mathbf{Q} \otimes \mathbf{Q}]^\dagger \bullet \mathbf{Q} \tag{31}$$

where $\dagger$ denotes pseudo-inverse.

It is evident from Equation 31 that Right Cauchy-Green deformation tensor, $\mathbf{C}$, is a function of the length of the edges, $d_{mn}^e$. Hence, so are all elements of $\mathbf{I}_V$ and therefore, so is the hyper-elastic strain energy, $\mathbf{\Psi}^e$. □

### 3.6 Updated System Equations

Now that we have proved that the updated hyper-elastic strain energy can be represented as a function of the edge lengths of the mesh, we rewrite the FEM system dynamics of the mesh to incorporate fracture. The parameters of the system dynamics of fractured object are updated based on suggestions from [40].

Internal elastic force is updated as, using Equations 7 and 9,

$$\begin{aligned} \mathbf{f}_e^{int^*} &= \chi^e \int_{\Delta_e} \frac{\partial \mathbf{\Psi}^e}{\partial \mathbf{u}} d\xi = \chi^e \int_{\Delta_e} \left(\frac{\partial \mathbf{\Psi}^e}{\partial \mathbf{F}}\right) \left(\frac{\partial \mathbf{F}}{\partial \mathbf{u}}\right) d\xi \\ &= \chi^e V_e \mathbf{P}(\mathbf{F}) \left(\frac{\partial \mathbf{F}}{\partial \mathbf{u}}\right) = \chi^e \mathbf{f}_e^{int} \end{aligned} \tag{32}$$

Then, from the definition of the stiffness matrix for an individual element in Equation 10

$$\begin{aligned} \mathbf{k}_e^* &= \chi^e \int_{\Delta_e} \frac{\partial \mathbf{f}_e}{\partial \mathbf{u}} d\xi = \chi^e \int_{\Delta_e} \left(\frac{\partial \mathbf{f}_e}{\partial \mathbf{F}}\right) \left(\frac{\partial \mathbf{F}}{\partial \mathbf{u}}\right) d\xi \\ &= \chi^e V_e \left(\frac{\partial \mathbf{F}}{\partial \mathbf{u}}\right)^T \left(\frac{\partial \mathbf{P}}{\partial \mathbf{F}}\right) \left(\frac{\partial \mathbf{F}}{\partial \mathbf{u}}\right) = \chi^e \mathbf{k}_e \end{aligned} \tag{33}$$

Stiffness matrix of full system can be written from Equation 11 as,

$$\mathbf{K}^* = \sum_{e=1}^{n_{tet}} \chi^e \mathbf{k}_e \tag{34}$$

The mass matrix remains same as Equation 5 and the damping matrix takes the form of $\mathbf{b}_e^* = \alpha \mathbf{m}_e + \beta \mathbf{k}_e^* + \tilde{\mathbf{b}}_e$.

## 4 IMPLEMENTATION

Plugging the updated parameters into Equation 4, we solve it for the complete fracture simulation. In order to implement our model for performing quasi-static simulations with no mass or damping matrix for regularization, we extend the code base released by [36]. For the complete dynamic simulation, we implement our model using the Vega FEM library [37].

Next we present details of how we create visualizations of the fracture simulation. Following that, we discuss the collision detection for fractured pieces. Then we present a complete algorithm for our model of fracture simulation.

### 4.1 Surface Remeshing for Visualization

Our FEM computations are performed on the graph induced by the computational mesh, that is never remeshed. However, in order to visualize the fracture we need to split the mesh used for visualization. We clarify this distinction between the computational and visualization meshes using a simple 2D example. Figure 7b depicts that when a fracture occurs, the system dynamics is evaluated on the computational mesh with weakened elements. When we need to render the fractured elements, we split the mesh used for visualization, as shown in Figure 7c and reconstruct the fracture surface. The reconstruction of the fracture surface for visualization adds very little computational overhead to the overall simulation cost. We have already explained how the system dynamics is computed on the computational mesh in the presence of fracture.

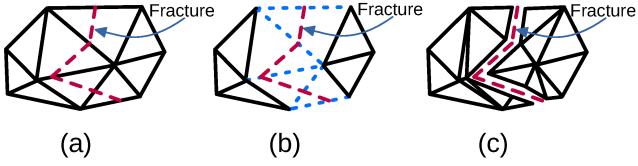

Fig. 7: (a) Original mesh with a fracture, (b) Computational mesh, with damaged edges marked in blue, on which system dynamics get evaluated, (c) Mesh for visualization is split and the fracture surface is reconstructed.

We now explain the details of how we split the visualization mesh for rendering. We use a node associated four-split configuration of a tetrahedron, as shown in Figure 6c, for fracture visualization. When a tetrahedral element splits, the corresponding edges and faces of the element split too. As illustrated in Figure 6, in our current work we always split edges, faces and tetrahedra with respect to their centroids.

When normal stress along an edge, *PE1PE2* (Figure 6a), crosses the critical threshold, it splits into two segments *PE1SE1* and *PE2SE2*. *PE1* and *PE2* are original nodes in the mesh. Let us call them parent nodes. The other two newly generated nodes, *SE1* and *SE2*, are called child nodes to *PE1* and *PE2*, respectively. The child nodes do not contribute to the system dynamics but rather follow movements of their parent nodes. In a similar manner, when a face gets fractured into three segments (Figure 6b), three child nodes, *SF1*, *SF2* and *SF3* are assigned to their corresponding parent nodes, *PF1*, *PF2* and *PF3*. Similarly four child nodes, *ST1*, *ST2*, *ST3* and *ST4* (Figure 6c) are assigned to four parent nodes *PT1*, *PT2*, *PT3* and *PT4* for a damaged tetrahedron. It is important to note that a triangular face or tetrahedron may not split into all three or four segments during the simulation. For example, with reference to Figure 6b, if we assume that only both the segments containing node *PF1* get damaged i.e., edge *PE2PE3* remains undamaged, then in such a case, child node *SF1* follows the movement of parent node *PE1* but child node *SF2* and child node *SF3* follow the average movement of parent nodes *PE2* and *PE3*. When similar scenarios, although more complex, arise for a tetrahedron, we apply the same principle. All possible cases of different kinds of fracture are taken into account in our simulation.

### 4.2 Collision Detection

Note that after fracture the disjoint vertices has no internal elastic force acting on them. But external body forces e.g., gravity or impulse force, continue to be applied on all the vertices whether disjoint or not. So we need no special treatment for collision detection. Continuous Collision Detection [41] [42] is performed between the vertices of the computational mesh and spherical collider (see Figure 17, 18). For fractured pieces colliding with the floor, we use Discrete Collision Detection [43] routine (see Figure 18). However, currently we do not handle self-collision between the fractured pieces in our work.

### 4.3 Algorithm

---

**Algorithm 1:** Remeshing-Free Graph-based Fracture

---

Initialize FEM simulation;
Let $n_v$ be total no of vertices of computational mesh;
Let $[x]_{n_v \times 1}$ be initial position vector;
**while** *True* **do**
  **for** *Each element in computational mesh* **do**
    Calculate stress along the edges $\sigma_{mn}$;
    **if** $\sigma_{mn} > \sigma_{thres}$ **then**
      Fracture the edge;
      Remesh the surface of the corresponding visualization mesh;
    **end**
    Resolve all collisions with the computational mesh;
    Calculate impulse force due to collision;
    Add all external forces to the computational mesh vertices;
  **end**
  Build full system $[M]_{n_v \times n_v} [v]_{n_v \times 1} = [f]_{n_v \times 1}$;
  Solve for velocity vector $[v]_{n_v \times 1}$;
  **for** *Each vertex in computational and visualization meshes* **do**
    Update position vector by $[x]_{n_v \times 1} += \Delta t \cdot [v]_{n_v \times 1}$;
  **end**
**end**

---

In the implementation of this algorithm, for solving the system dynamics we use a regular Newton solver augmented with a line search [36] for quasi-static simulation and an implicit backward Euler integrator with a conjugate gradient solver [37] for dynamic simulation. All the simulations are performed on an Intel Core i7-9750H CPU with 12 threads at 2.60 GHz equipped with NVIDIA GeForce RTX 2060 graphics card. For visualization, the simulation results are raytraced in Houdini. Apart from the models obtained from the Stanford 3D scanning repository, the other 3D models used in this project are obtained from `free3d.com` and `lifesciencedb.jp`. Open-source software TetWild [44] and TetGen [45] are used to generate the volumetric simulation meshes.

## 5 RESULTS

We present multiple visualizations of fracture simulation for a variety of materials to demonstrate the robustness of our method and the diversity it can handle. We present quasi-static fracture simulation results first to exhibit the high stability achievable with our algorithm. Next we show dynamic fracture simulation results which mimic real world events like collision of hard objects with solid jade or glass objects and glass shells. We also show the effect of applying

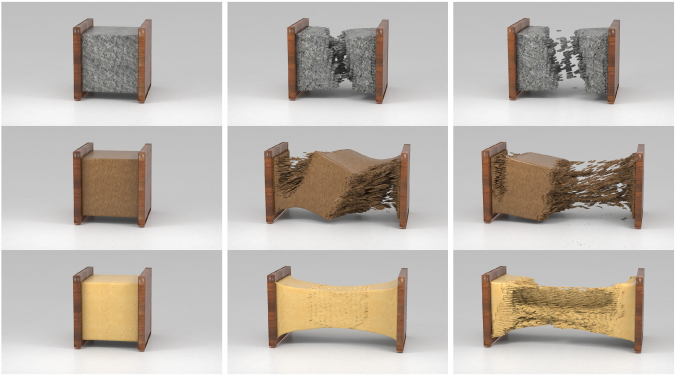

Fig. 8: Blocks made of different material are pulled apart in a quasi-static fracture simulation. We simulate blocks of stone (top row), soft Balsa wood (middle row) and cheese (bottom row).

tensile or impact forces to rubber-like, and jello-like materials.

Next, we experimentally validate our model by recreating a few benchmark fracture experiments in simulation that are commonly performed in a structural mechanics laboratory, with our framework and compare the results with expected results from similar experiments performed in the real world. These experimental simulations are complete dynamics simulations that demonstrate the accuracy of our method. Finally, we compare our method with existing related works.

Mesh resolution and timing results for the simulations are reported in Table 2.

| *Simulation* | # *Tetrahedra* | *sec/frame* |
|---|---|---|
| Stone (Fig. 8, 9) | 168.2k | 178.1 |
| Soft Balsa wood (Fig. 8  9) | 168.2k | 189.6 |
| Cheese (Fig. 8, 9) | 168.2k | 214.5 |
| Loaf (Fig. 1) | 620.6k | 291.7 |
| Jello armadillo (Fig. 2) | 159.1k | 78.9 |
| Femur bone (Fig. 10) | 21.6k | 9.3 |
| Solid glass armadillo (Fig. 17) | 4.7k | 2.1 |
| Shell glass armadillo (Fig. 17) | 41.1k | 20.7 |
| Solid tempered glass bunny (Fig. 18) | 23.7k | 11.8 |
| Shell tempered glass bunny (Fig. 18) | 38.8k | 17.2 |
| Rubber bunny (Fig. 11) | 23.7k | 10.9 |
| NVE 36 steel bar (Fig. 12, 13) | 27.8k | 13.4 |
| AWS 7018-G steel bar (Fig. 12, 13) | 27.8k | 13.1 |
| Cement cylinder (Fig. 14) | 45.8k | 24.6 |

TABLE 2: Model resolution and the time taken per frame for all the simulations.

## 5.1 Quasi-Static Simulations

Here we use the stable Neo-Hookean energy defined in [36].

$$\mathbf{\Psi}_{neo} = \frac{\mu}{2}\left(I_C - 3\right) + \frac{\lambda}{2}\left(J - \alpha\right) - \frac{\mu}{2}\log\left(I_C + 1\right) \quad (35)$$

where $\mu$ and $\lambda$ are Lamé parameters and $\alpha = 1 + \frac{\mu}{\lambda} - \frac{\mu}{4\lambda}$.

Quasi-static experiments are performed in the absence of gravity, and mass & damping matrix regularizers. Using Equations 34, 1, 10 and 32, we can formulate the a quasi-static simulation as

$$\sum_{e=1}^{n_{tet}} \chi^e \mathbf{k}_e \mathbf{u} = \sum_{e=1}^{n_{tet}} \chi^e \mathbf{f}_e^{int} \quad (36)$$

First we take rectangular blocks of three materials: stone, soft Balsa wood and cheese whose points at both ends are fixed. Then we pull each block from both ends by changing position of fixed points and examine the progressive fracture that results. This is shown in Figure 8 column wise, from left to right. Fracture of stone, soft Balsa wood and cheese are illustrated row wise from top to bottom. It can be seen from the figure that maximum diffusion of cracks is present in cheese and minimum in stone. We control the extent of crack diffusion in our method using different sizes of kernel support, $R_d$, (see Equation 20) for different material. Here e.g., we use $R_d = 0.0$ for stone, $R_d = 0.1$ for wood and $R_d = 0.5$ for cheese. This supports the claim that diffusion of cracks increase with increasing support of kernel. In Figure 9 we simulate the twisting of the same material blocks from both ends. While the stone block fractures instantly, the soft Balsa wood block breaks apart after much twisting. The block made of even softer cheese like material does not break on twisting.

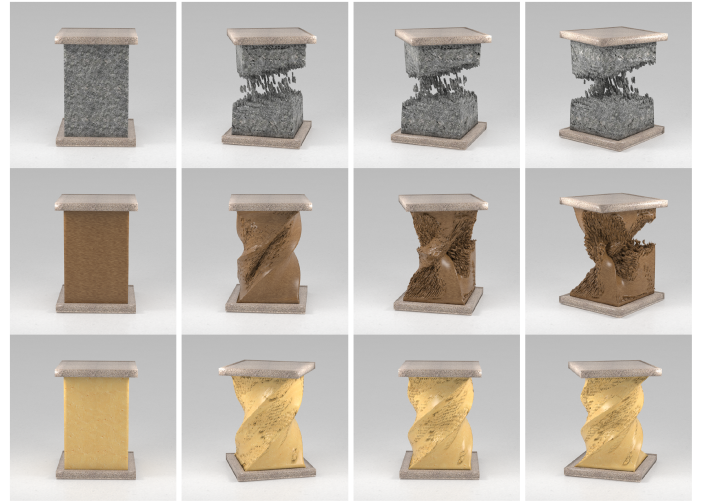

Fig. 9: Blocks made of different material are twisted in a quasi-static fracture simulation. We simulate blocks of stone (top row), soft Balsa wood (middle row) and cheese (bottom row).

The quasi-static experiments exhibit the high stability of our method. Notice that in Figure 8, 9 the fractured pieces are hanging in the midair in the absence of gravity. Their residual velocities do not get blew up and thus, these fractured pieces do not move away or diverge. Moreover, the quasi-static experiments give us a proper understanding about the effect of crack diffusion controlling parameter on fracture simulation. Along with Young's modulus and Poisson's ratio, this is also a very important tool for simulating fracture of different kinds of materials.

## 5.2 Dynamic Simulations

In a complete dynamic simulation we take into account the effects of external forces and solve full system dynamics as presented in Equation 2 with mass & damping matrix regularizers.

Here we use the invertible Saint–Venant Kirchhoff energy [37] defined as below.

$$\Psi_{stvk} = \frac{\lambda}{8}\left(I_C - 3\right)^2 + \frac{\mu}{4}\left(II_C - 2I_C + 3\right) \tag{37}$$

where $\mu$ and $\lambda$ are Lamé parameters. To begin with, in

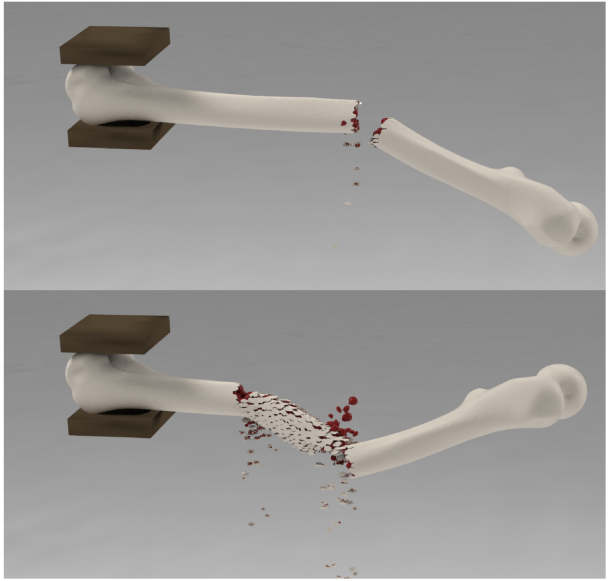

Fig. 10: Fracture of bone: splitting (top) and buckling (bottom).

Figure 1 we illustrate the highly detailed and intricate fracture effects produced by tearing a loaf of bread. It is evident from these figures that our simulation model can model complex fracture patterns and can easily scale to high resolution meshes. In Figure 2 we simulate the limbs of a jello armadillo being ripped off. Here we also visualize the corresponding strain profiles. In the frame shown on the leftmost side the armadillo appears just prior to fracture. Thus, it has highest strain at elements that are going to be damaged (coloured red). Once damage sets in post fracture, as shown on the right, the strain decreases again. The disconnected nodes in the damaged mesh have no strain value and thus appear in blue-green. In Figure 10, we show frames from two simulations of fracture in a femur bone. Bone displays both brittle and quasi-ductile behavior depending on the direction of loading being applied [46]. The top image shows a simulation where we hold the bone at one end and apply a side impact force, thus causing a brittle, transverse split fracture [47], caused by shear stresses that develop in the bone. On the bottom, we compress the bone, causing it to bend and behave in a quasi-ductile manner, producing a butterfly, comminuted or open fracture [47]. This behavior of the bone emerges automatically from our simulation model.

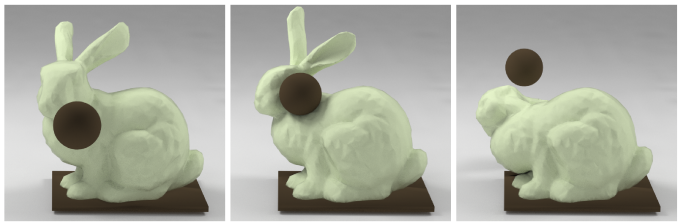

Fig. 11: Deformation of a rubber bunny without any fracture.

In the top row of Figure 17, a solid ball hits an armadillo made of solid blue jade to generate large chunks of debris from the fracture. In the bottom row of the same figure, a solid ball smashes through a glass shell model of armadillo, leaving a hole in the model with associated debris.

When tempered glass, which is used in a car windscreen gets fractured it leaves a lot of spraying debris as if the glass has exploded. Using our fracture simulation method, in the top row of Figure 18 we render the fracture of a bunny made of solid tempered glass when it is hit by a solid ball. The fracture of a shell model of a bunny made of tempered glass is simulated in the bottom row of the same figure. Notice the intricate fracture pattern and spraying debris generated in both cases.

To depict that our method is capable of rendering proper deformation without any fracture, we simulate and visualize a ball hitting a rubber bunny in Figure 11. The model bends on impact, however, does not break.

### 5.3 Experimental Validation

In order to evaluate and validate how faithfully real world fracture can be simulated using our method, we present a qualitative and quantitative comparison with three benchmark laboratory fracture experiments. These are the Charpy impact test, the Izod impact test and the Brazilian test.

#### 5.3.1 Charpy Impact Test

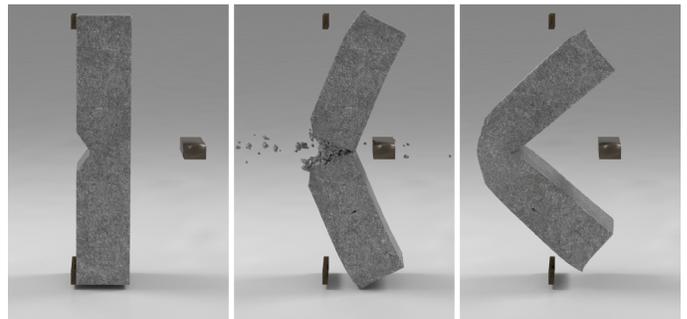

Fig. 12: **Charpy Impact Test**: The left image shows the configuration of the Charpy test. In the middle image a specimen with low tensile strength splits into two, while on the right a specimen with high tensile strength only bends and gets partially damaged.

In this test we use a steel specimen of dimension $10mm \times 10mm \times 55mm$ with a $45°$ 'V' shaped groove in the middle of it (see left image of Figure 12). A pendulum of known mass and length then impacts the back of the groove with both ends of the specimen held against fixed beams. Now depending on the tensile strength of the material, the specimen breaks fully into two pieces or is partially damaged and bends in a three point configuration. We simulate this experiment using our method, on specimens made of two different kinds of steel with different tensile strengths (the name of the steel is mentioned in Table 2). Instead of the entire swinging pendulum, as present in the real Charpy test setup, we simulate an impact with a fast moving block (as can be seen in the image) that hits the specimen at the same location where the specimen is supposed to be hit in the real test. As shown in Figure 12 the steel with lower tensile strength (middle image) breaks into two pieces while the other one (right image) splits partially and bends. Simulation material parameters used in this experiment are obtained from [48].

### 5.3.2 Izod Impact Test

The Izod impact test specimen is shown in the left image of Figure 13. Here the specimen is held in a cantilevered beam configuration with one end fixed and the pendulum hits it on the other end. The results of Izod test performed on the same materials as before, are shown in middle and right images of the figure.

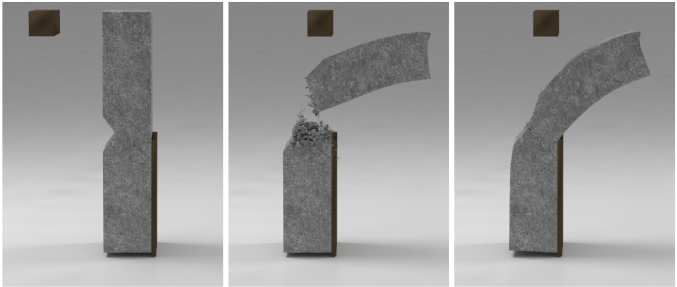

Fig. 13: **Izod Impact Test**:The left image shows the configuration of the Izod test. In the middle image a specimen with low tensile strength splits into two, while on the right a specimen with high tensile strength only bends and gets partially damaged.

In Figure 15 we plot the load-displacement curves for the Charpy (middle) and Izod (right) tests respectively as obtained from our simulated experiments. The green curve denotes brittle fracture, which splits sharply beyond a load threshold. Red curve which denotes ductile fracture, indicates a plastic flow in the material. The brittle and ductile fracture curves for both tests closely resemble the theoretical force-displacement curve shown on the left in the same figure [48]. Brittle material fractures completely after elastic region whereas ductile material partially fractures and then bends due to plasticity. This proves that brittle and ductile material simulation performed by our method closely matches the behaviour of real world materials, both qualitatively and quantitatively.

### 5.3.3 Brazilian Test

The Brazilian test is a laboratory test for indirect measurement of tensile strength. In this test a vertical load is applied at the highest point of a cylinder, the axis of which is placed horizontally, and is supported on a horizontal plane. Here we use a cylinder of diameter $100mm$ and length $255mm$ made of ordinary portland cement. In Figure 14 the initial (left) and fractured (right) configuration of the cylinder are shown. We compare the precision of the experiment performed in simulation using our method with similar real experiment from literature [49] by plotting their corresponding load displacement curves in Figure 16. The close match between the two curves validates the accuracy of our simulation model.

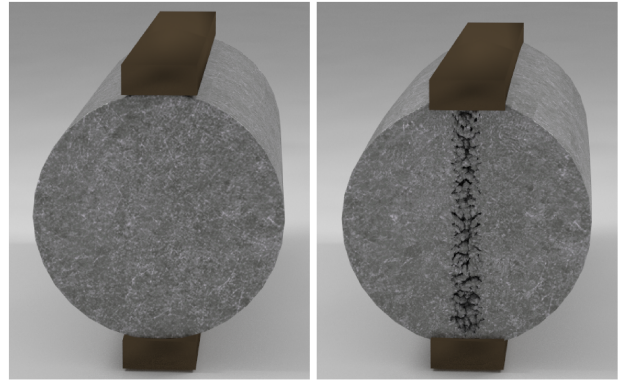

Fig. 14: **Brazilian test**: Initial (left) and fractured (right) configuration of cement cylinder.

## 5.4 Comparison with Existing Techniques

To the best of our knowledge, this is the first work in computer graphics that presents a FEM-based fracture method that is graph-based, remeshing-free, highly stable and can incorporate large number of cracks with little extra computational overhead on FEM. Unlike XFEM [4] or VNA [28] [29], the size of system matrix remains constant throughout the simulation, thus reducing the computational cost over standard FEM substantially. While a state of the art XFEM-based fracture simulation [4] requires 1539 sec to render a single frame for a 40k model, our model requires 291.7 sec to render a frame for the simulation of an extremely complex loaf model consisting of 620.6k tetrahedra. Moreover, simulating spraying debris like fracture of tempered glass with existing FEM or XFEM based solutions in literature will require prohibitively expensive numerical computation due to remeshing and rapid scaling of the system matrix. Our method can simulate a very high amount of fracture debris easily. Our method enjoys the good characteristics of both FEM and XFEM methods, without the overheads. Additionally, it does not suffer from the limitations of MPM and BEM based methods such as difficulty in enforcing essential boundary conditions, poor performance in rendering fracture of rigid objects, rigidity of fragmented elements and expensive computation. A feature based comparison of our model over other existing frameworks is presented in Table 3.

## 5.5 Difference Between Graph-Based FEM and Mass-Spring Model

It can be tempting to assume that our model is similar to mass-spring or peridynamics based fracture model [11], reformulated as FEM. However, there exists two crucial differences.

First, in peridynamics or mass-spring model the total energy an element with $n_e$ number of nodes, can be formulated as

$$\boldsymbol{\Psi}^e = \sum_{i=1}^{n_e}\sum_{i>j}^{n_e} \boldsymbol{\Psi}^e_{ij} \qquad (38)$$

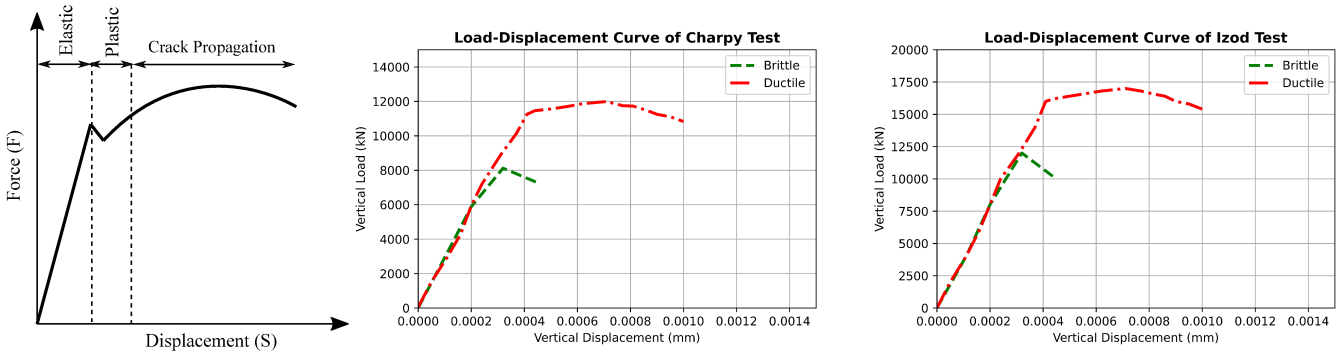

Fig. 15: The left plot shows the theoretically expected force-displacement curve for the fracture of a material. The center and the right plots show the load-displacement curve of our simulated Charpy and Izod impact tests respectively.

| Comparison Table | | | | | | |
| --- | --- | --- | --- | --- | --- | --- |
| *Method* | *R.F.* | *R.C.S.* | *E.E.B.* | *C.C.* | *R.D.B.* | *Stability* |
| FEM [1] [25] [26] | No | No | Easy | High | Yes | High |
| XFEM [4] | Yes | Yes | Easy | High | No | Low |
| VNA [28] [29] | Yes | Yes | Easy | High | Yes | Low |
| BEM [12] [13] | No | No | Easy | Very High | No | High |
| Peridynamics [10] [11] | Yes | No | Easy | Very High | Yes | Low |
| MPM [8] [9] | Yes | No | Difficult | Medium | Yes | High |
| Our method | Yes | No | Easy | Very Low | Yes | Very High |

TABLE 3: Comparison with existing frameworks. R.F., R.C.S., E.E.B., C.C. and R.D.B. denote Remeshing-Free, Require Changing the Initial System matrix, Enforcing Essential Boundary conditions, Computational Cost and Render both Ductile and Brittle Fracture respectively.

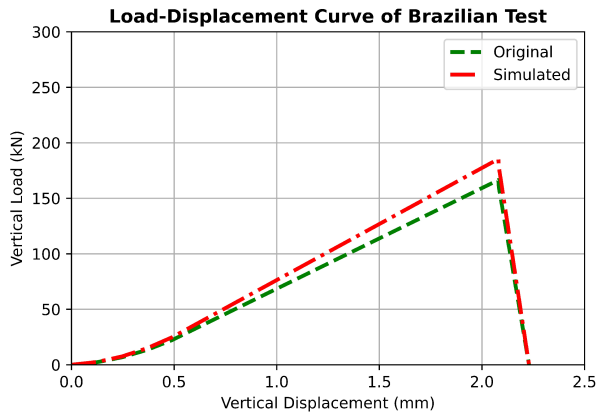

Fig. 16: **Brazilian test**: Load displacement curve for original and simulated experiments.

That is, the strain energy is just the sum of the strain energies along each edge without any cross dependence. But in case of graph-based FEM method, strain energy depends on the distance between all the edges of the element which is not just a simple sum. This is also observed by [14].

Second, in case of peridynamics, force at any point depends on points far away from it. Therefore it is a completely non-local dynamics, whereas in graph-based FEM, force values of an element are derived locally from the same element stresses.

## 6 CONCLUSION

We present a novel graph-based remeshing-free FEM approach for ductile and brittle fracture. We derive a theoretical proof to show that our method extends to non-linear hyper-elastic strain energies. We follow this with the complete algorithmic description of our model. The high stability, speed and robustness of our method is illustrated and established via both quasi-static and dynamic experiments. We evaluate the appeal and realism of our fracture simulations through results on objects made of a wide variety of materials. Comparing with benchmark fracture experiments, we validate the correctness and accuracy of our method.

### 6.1 Limitations and Future Work

Even though we believe that our method is capable of producing particularly realistic fracture of a variety of materials at a high speed, it does have some limitations in it's current form. We do not explore anisotropic fracture and viscoplasticity in this work. Also, currently an edge of an element in the object mesh can split in maximum two parts, thus limiting the fracture of tetrahedral elements at maximum to four parts. We wish to extend our work to incorporate high number of fracture segments in a single element.

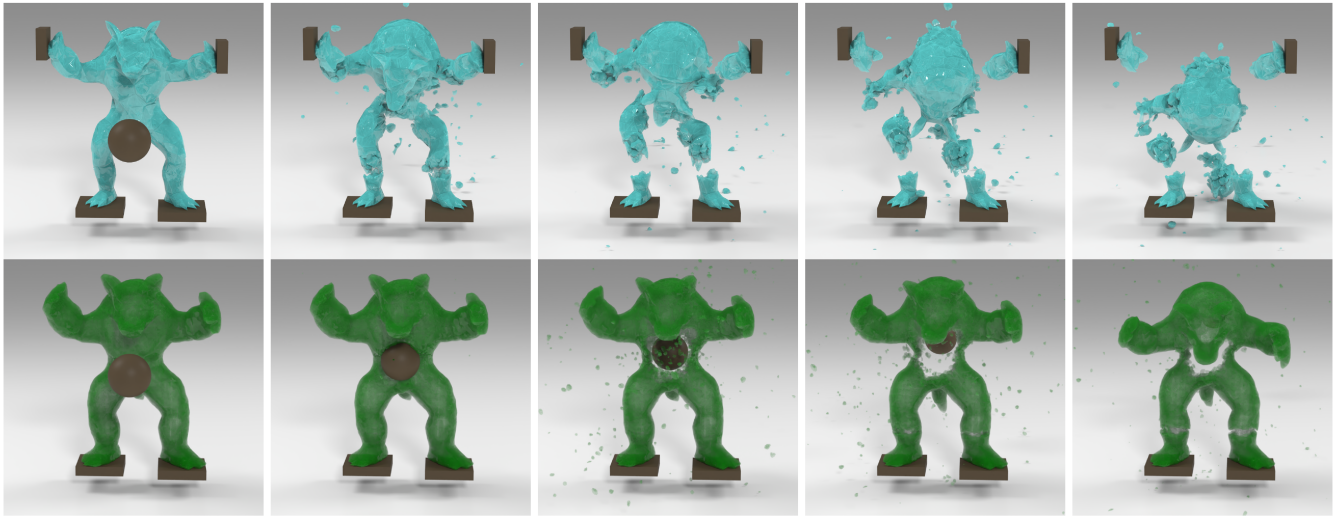

Fig. 17: Fracture of an armadillo model made of solid blue jade (top row) and glass shell (bottom row).

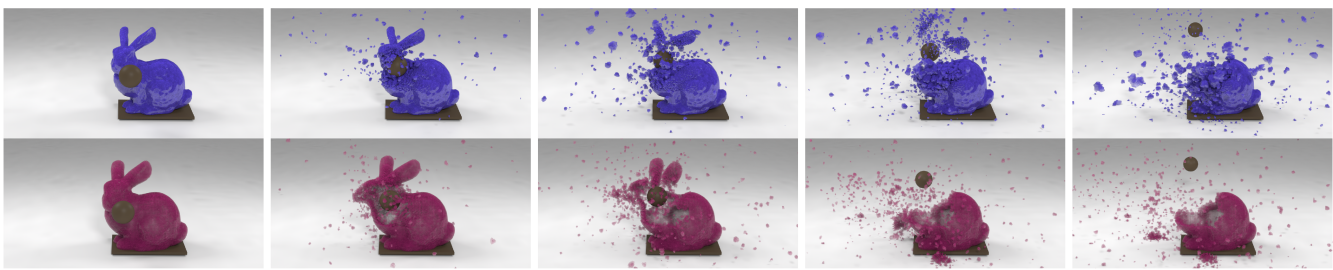

Fig. 18: Fracture of a bunny model made of solid tempered glass (top row) and a tempered glass shell (bottom row).

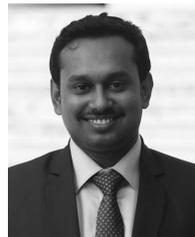

**Avirup Mandal** received a B.E. in Electrical Engineering from Jadavpur University, in 2015. He is working toward the MTech+PhD dual degree at Indian Institute of Technology Bombay. His research interests include physics-based animation, computer graphics, fracture dynamics and computational haptics.

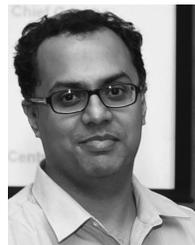

**Parag Chaudhuri** is an associate professor in the Department of Computer Science and Engineering at the Indian Institute of Technology Bombay. He holds a B.E. in Civil Engineering from the Delhi College of Engineering and a Ph.D. from the Department of Computer Science and Engineering at the Indian Institute of Technology Delhi. He has worked as a postdoctoral researcher at MIRALab, University of Geneva. His current research interests broadly cover the areas of computer graphics, virtual and augmented reality and computer vision. His current work centers around the animation of characters and natural phenomena in virtual worlds.

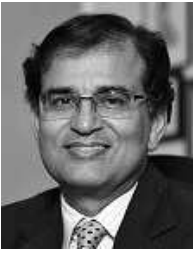
**Subhasis Chaudhuri** is currently Director of IIT Bombay and K.N. Bajaj Chair Professor, Department of Electrical Engineering. He did his B.Tech. in Electronics & Communication Engineering from IIT Kharagpur in 1985 and M.Sc. in Electrical Engineering from University of Calgary in 1987 and Ph.D. in Electrical Engineering in 1990 from University of California, San Diego. He has received several major awards recognizing his excellence in research, which include the Shanti Swarup Bhatnagar Prize, Swarnajayanti Fellowship, J.C. Bose Fellowship and the Vikram Sarabhai Award among several others. He was conferred with the prestigious title "TUM Ambassador" by Technical University Munich (TUM), Germany. He is a Fellow of Indian National Science Academy (INSA), Indian Academy of Sciences (IASc.), Indian National Academy of Engineering (INAE), National Academy of Science India (NASI) and IEEE, USA. He has been recognized as a Distinguished Alumnus of IIT Kharagpur. He has held editorial positions in various journals like IEEE PAMI, IJCV and SIIMS. He has also served as the technical program co-chair for ICCV. His primary areas of research include pattern recognition, computer vision and computational haptics.